\documentclass[journal=jacsat,manuscript=article]{achemso}

\usepackage[version=3]{mhchem} 
\usepackage{upgreek}
\usepackage{amssymb}

\author{Zhichao Li}
\affiliation{Department of Electrical and Computer Engineering, Rice University, Houston, TX 77005, USA}
\alsoaffiliation{Applied Physics Program, Rice University, Houston, TX 77005, USA}

\author{Ciril S. Prasad}
\affiliation{Department of Electrical and Computer Engineering, Rice University, Houston, TX 77005, USA}
\alsoaffiliation{Applied Physics Program, Rice University, Houston, TX 77005, USA}

\author{Xielin Wang}
\affiliation{Department of Electrical and Computer Engineering, Rice University, Houston, TX 77005, USA}
\alsoaffiliation{Applied Physics Program, Rice University, Houston, TX 77005, USA}

\author{Ding Zhang}
\affiliation{Department of Physics and Astronomy, Rice University, Houston, TX 77005, USA}

\author{Gururaj V. Naik}
\email{guru@rice.edu}
\affiliation[Rice University]
{Department of Electrical and Computer Engineering, Rice University, Houston, TX 77005, USA}

\title[]
  {Sensing beyond the exceptional point for high detectivity}

\abbreviations{IR,NMR,UV}
\keywords{American Chemical Society, \LaTeX}

\begin{document}







\begin{abstract}

Exceptional point (EP)-based optical sensors exhibit exceptional sensitivity but poor detectivity due to their acute sensitivity to perturbations such as noise. When the optical budget is limited as in applications on mobile platforms, high detectivity might be equally important as high sensitivity. In such scenarios, off-EP sensing is advantageous where a slight loss in sensitivity can lead to a significant gain in detectivity. Here, we show that a passive parity-time symmetric plasmonic-photonic hybrid shows a peak in its detectivity off EP where the sensitivity is still very high -- surpassing the sensitivities of equivalent fully-plasmonic and fully-photonic systems. We demonstrate the high detectivity of off-EP sensing of anti-mouse IgG protein using this plasmonic-photonic hybrid system. We report a sensitivity of 1.2 nm$/$nM while requiring a minimum optical budget of 1.3 nJ. Thus, non-Hermitian plasmonic-photonic hybrids could be the best class of nanophotonic sensors for real-world applications.
\end{abstract}


The Exceptional Points (EPs) of non-Hermitian 
optical devices have garnered much attention, especially in sensing applications, for their exceptional sensitivity \cite{peng2014parity,alaeian2014parity,wiersig2014enhancing,wiersig2016sensors,chen2017exceptional,hodaei2017enhanced,el2019dawn,chen2019sensitivity,dong2019sensitive,djorwe2019exceptional,hokmabadi2019non,ozdemir2019parity,miri2019exceptional,wang2020petermann,park2020symmetry,zeng2021ultra}. However, the exceptional sensitivity at EP has also led to debates about the detectivity of these sensors \cite{lau2018fundamental,langbein2018no,smith2019parity,horstman2020exceptional,smith2022beyond}. Detectivity is the sensor's ability to distinguish signal from noise \cite{nudelman1962detectivity}. The exceptional sensitivity of EP-based sensors has been shown to result in poor detectivity making them practically useless, especially in limited energy budget scenarios \cite{horstman2020exceptional,smith2022beyond}. One possible solution to this problem is to exploit the inherent trade-off between detectivity and sensitivity 
 in nanophotonic sensors \cite{barton2021high,li2023balancing}. Operating the sensor a little far from the EP reduces sensitivity, but could significantly boost the detectivity. To understand this point, here, we study the detectivity, sensitivity, and Q-factors of a passive parity-time (PT) symmetric \cite{bender1998real,heiss2004exceptional,heiss2012physics} sensor operating near its EP. 

\begin{figure}[htpb]
    \centering
    \includegraphics[width=1\textwidth]{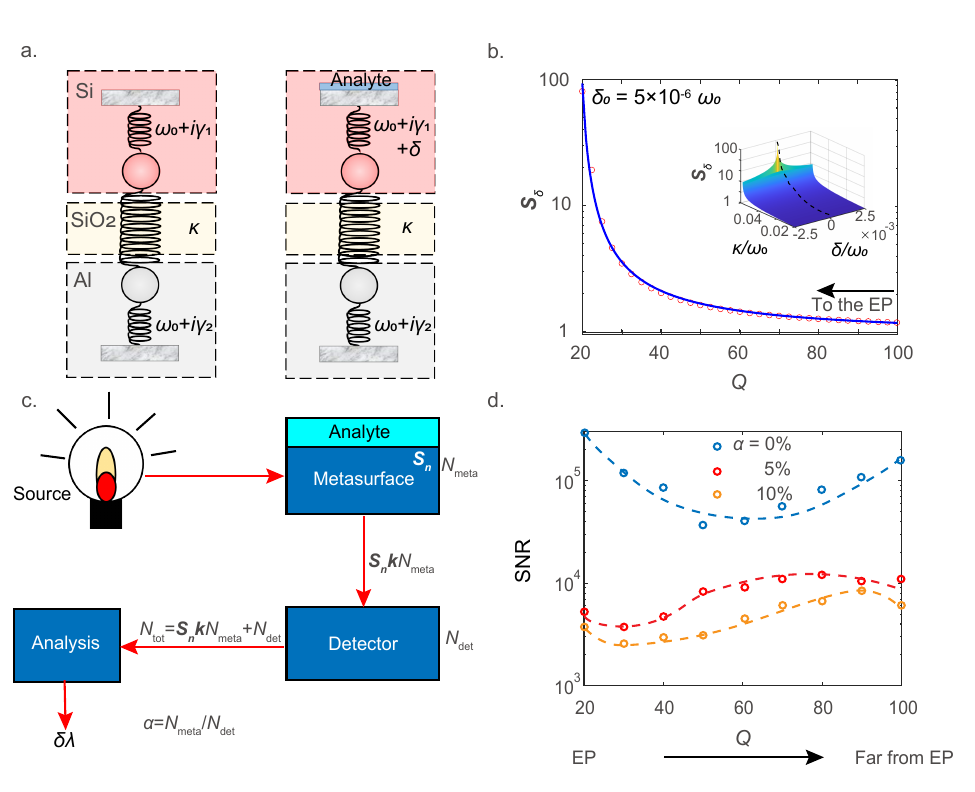}
    \caption{Sensitivity vs. detectivity in non-Hermitian systems: a. A schematic spring-mass model of a passive parity-time (PT) symmetric system for sensing b. The trade-off between sensitivity and Q-factor. The inset shows the trade-off landscape as a function of perturbations to coupling constant $\kappa$ and detuning $\delta$. The dashed line in the inset is shown in the main plot. Red circles correspond to the simulation points and the blue line is the best-fit curve. c. The noise model of the non-Hermitian sensor: Noise from the sensing metasurface $N_{\rm meta}$ is amplified by its sensitivity, and adds up to the detector noise $N_{\rm det}$, and then, transforms in the analysis unit to appear as the noise in the peak shift estimate. d. Signal-to-noise ratio (SNR) of the non-Hermitian system for three different values of $\alpha = N_{\rm meta}/N_{det}$. Detectivity peaks off EP when $\alpha \ne 0$.}
    \label{fig1}
\end{figure}

Our passive PT-symmetric sensor employs an array of Si-disk resonators to support nearly lossless photonic modes. We couple these photonic modes to their lossy images in an aluminum ground plane to create a passive PT-symmetric system. Fig.~\ref{fig1}a shows the model of our system. The coupling between the two modes $\kappa$ is tuned by varying the thickness of the silicon dioxide spacer layer between the top Si-disks and the bottom aluminum ground plane. The sensing action of this coupled resonator system emerges from the analyte interacting with the top silicon photonic resonator to detune its resonant frequency by $\delta$. A mathematical model of this system shows that the sensitivity of this sensor is highest at its EP where the Q-factor is smallest (Fig.~\ref{fig1}b). This minimum Q-factor and the corresponding maximum sensitivity are arbitrarily chosen in this study. Note that the sensitivity at EP is not infinite because our system is passive with net optical loss. As the coupling between the two resonators is gradually decreased from the EP condition, the system enters the broken symmetry regime where the Q-factor consistently rises. However, the sensitivity of this sensor consistently falls as the Q-factor increases. This trade-off between sensitivity and Q-factor may be captured by a coupled mode theory-based model to result in Eq.\ref{eqn:fitting}. Fig.\ref{fig1}b plots the sensitivity vs. Q-factor from Eq.\ref{eqn:fitting} as a blue curve. The red circles correspond to the direct solution from the coupled mode theory.

\begin{equation}
    \boldsymbol{S}_{\delta}  -1 + A = A \frac{Q_{\rm max} - Q_{\rm EP}}{Q - Q_{\rm EP}} \\
\label{eqn:fitting}
\end{equation}

\noindent where $A$ is a dimensionless constant that depends on the sensor parameters. Here, $\boldsymbol{S}_{\delta}$ is defined as the dimensionless sensitivity defined as the magnitude of the change in eigenvalue for a small detuning $\delta$ of the nearly lossless resonator divided by $\delta$.  A more common definition of sensitivity used in such studies corresponds to the magnitude of the change in eigenvalue for a small change in the refractive index of the analyte layer, denoted by $\boldsymbol{S}_{n}$. The two sensitivity definitions are related to each other by a dimensional proportionality constant. 

Next, we model the noise in this sensor system as shown in Fig.~\ref{fig1}c. Each part of the sensing system contributes its noise to the final result. The noise from the metasurface $N_{\rm meta}$, which includes mechano-optic and thermo-optic sources, is amplified by the sensitivity $\boldsymbol{S}_{n}$ of the sensor. The factor $k$ is a dimensional proportionality constant. Then the amplified metasurface noise adds up to the photodetector noise $N_{\rm det}$. Finally, the raw data containing the noise from all the components in the optical path is transformed in the analysis unit. The final sensing result -- the concentration of the analyte related to the changes in the resonance peak position and the resonance linewidth -- is thus mixed with the transformed version of the total system noise. In this work, we assume all the noise sources are additive white Gaussian sources and the raw data analysis is a Gaussian fit routine to find the resonance peak. Also, we define noise ratio $\alpha=\frac{N_{\rm meta}}{N_{\rm det}}$ which captures the effect of noise amplification by the sensor sensitivity on the overall noise. Using the error analysis for a least mean square Gaussian peak fitting procedure to find the resonance peak~\cite{lenz1992errors}, the signal-to-noise ratio (SNR) for a fixed optical energy budget $E$ (product of incident intensity and data acquisition time) in the extracted peak shift $\delta\lambda$ is \cite{li2023balancing} 

\begin{equation}
     {\rm SNR}_{\delta\lambda}= \frac{C}{\lambda_{0}} \sqrt{\frac{\rm FWHM}{d\lambda}} \frac{\Delta n \boldsymbol{S}_{n} Q}{\boldsymbol{S}_{n} k\alpha+1} \frac{E}{N_{\rm det}} = \frac{C}{\sqrt{\lambda_0 d\lambda}}  \frac{\Delta n \boldsymbol{S}_{n} \sqrt{Q}}{\boldsymbol{S}_{n} k\alpha+1} \frac{E}{N_{\rm det}}
    \label{eqn:det1}
\end{equation}

\noindent where $\lambda_0$ is the peak wavelength of the sensor with no analyte, $\Delta n$ is the change in the effective refractive index of the analyte layer, ${\rm FWHM}$ is the full width-half maximum of the resonance, $d\lambda$ is the spectral resolution of the spectrophotometer, and $C$ is a proportionality constant. Fig.~\ref{fig1}d plots the simulated SNR of the sensor for different $\alpha$ values in the broken symmetry regime. Simulations are carried out by adding 80 dB detector noise to the resonance spectra and carrying out multiple fittings to extract the resonance peak and linewidth. The lineshape of the resonance spectra is assumed Gaussian. When $\alpha$ is zero, the sensitivity of the sensor has no noise to amplify, and hence, EP is the best point of operation. However, even for a small $\alpha$ value of 5\%, the situation is entirely different. The highest SNR operation requires an off-EP sensor. The larger the value of $\alpha$, the farther from EP is the maximum SNR condition.

Using the definition of specific detectivity $D^*_{\rm det}$ of a detector (the reciprocal of noise equivalent power (NEP) per unit area of the detector), and the equation for SNR in Eq.~\ref{eqn:det1} the overall specific detectivity ($D^*$) of the optical sensing system is given by Eq.~\ref{eqn:det2}.

\begin{equation}
    D^*= \frac{C}{\sqrt{\lambda_0 d\lambda}}  \frac{\Delta n \boldsymbol{S}_{n} \sqrt{Q}}{\boldsymbol{S}_{n} k\alpha+1} D^*_{\rm det}
    \label{eqn:det2}
\end{equation}

\noindent where $C'$ is a proportionality constant. Given the same functional forms of detectivity and SNR equations, the sensors designed at the EP are not always the best \cite{vollmer2012review,luchansky2012high}. Off-EP sensors are promising for real-world applications.

\section{Results and discussion}
\begin{figure}[htpb]
    \centering
    \includegraphics[width=1\textwidth]{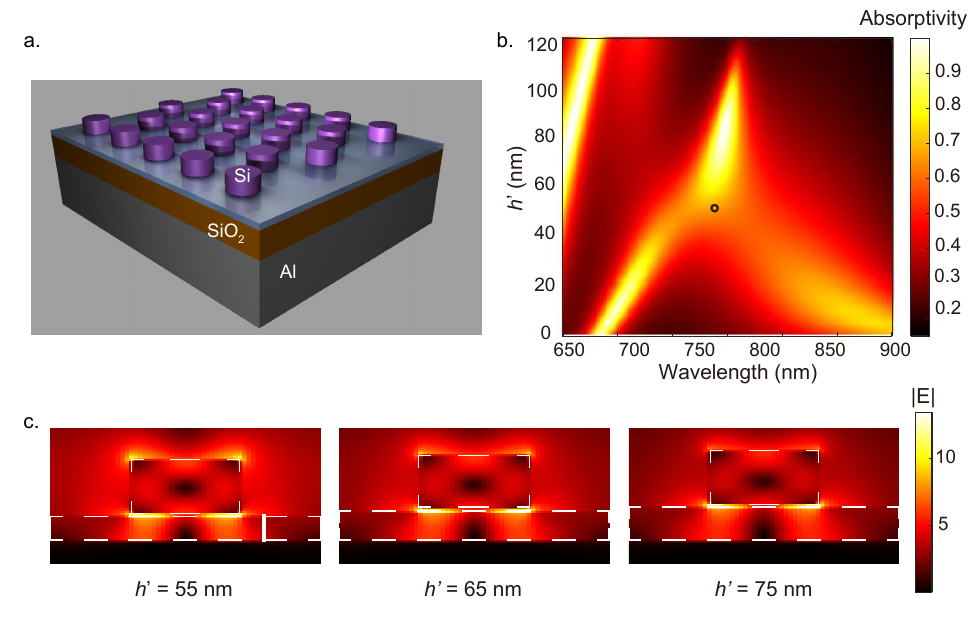}
    \caption{Design of the non-Hermitian plasmonic-photonic hybrid metasurface sensor: a. A schematic of the metasurface consisting of a hexagonal array of silicon cylinders with a period $P=$ 600 nm on top of a silica spacer layer. Aluminum is the substrate. The radius and height of the silicon cylinders are both 120 nm. The spacer thickness $h'$ is varied from 0 to 120 nm. The analyte interaction is modeled as a 5 nm-thick dielectric layer of refractive index between 1 and 1.001 sandwiched between the silicon and silica layers. b. The simulated absorption spectra of the structure with different spacer thickness $h'$. c. Field distribution in a cross-section of the structure at its resonance three chosen values of $h'=$ 55, 65, and 75 nm. The scale bar represents 50 nm.}
    \label{fig2}
\end{figure}

To demonstrate the high detectivity of off-EP sensors, we chose a passive PT-symmetric device comprising a hexagonal array of 120 nm-tall silicon cylinders on top of a silica spacer placed on a thick aluminum film. The height of the spacer $h'$ is varied from 0 to 120 nm. The distance between the two closest cylinders in the hexagonal array, henceforth called the period \textit{P}, was 600 nm. The schematic of the structure is shown in Fig.~\ref{fig2}a. Using a commercial finite-difference time-domain (FDTD) solver, we simulate the optical properties of our design under a normal incident plane wave light source. The absorption spectra for various spacer thicknesses are shown in Fig.~\ref{fig2}b. The silicon disk array supports a quasi-bound state in the continuum (q-BIC) mode which interacts with its image in the aluminum ground plane \cite{doiron2019non,yang2021non}. When $h'$ is small, the single q-BIC mode splits into two as expected for a strong coupling condition. Near the spacer thickness of 50 nm, the two modes coalesce into one to make an EP. When the spacer thickness is greater than 50 nm, the system works in the symmetry-broken regime or weak coupling regime. This weak coupling regime is the focus of this current work where the sensor operation is optimized.

The field distribution across the structure is shown in Fig.~\ref{fig2}c for three different spacer thicknesses. Very slightly off EP, $h'$ of 55 nm, the fields remain well-confined to the resonators. But, they spread more for thicker spacer conditions. Tight confinement of fields for thinner spacers results in high sensitivity. However, the better delocalization of the fields for thicker spacer results in smaller radiative loss, and hence higher Q-factors. The trade-off between sensitivity and Q-factor is inherent to all nanophotonic sensors \cite{li2023balancing}. However, if we compared the trade-off observed in our off-EP non-Hermitian system with those observed in entirely plasmonic or photonic systems, the non-Hermitian plasmonic-photonic hybrid offers many advantages over its entirely plasmonic or entirely photonic counterparts.

\begin{figure}[htpb]
    \centering
    \includegraphics[width=1\textwidth]{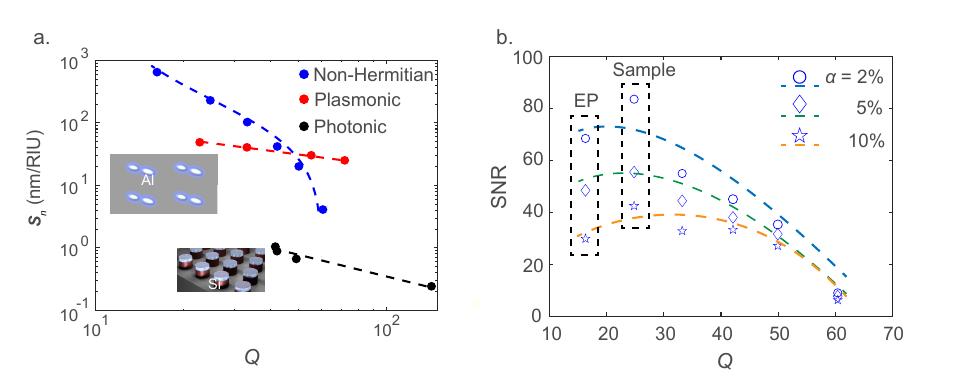}
    \caption{Sensitivity and SNR of the non-Hermitian plasmonic-photonic hybrid: a. The simulated sensitivity values for the non-Hermitian plasmonic-photonic hybrid, aluminum plasmonic dimer, and silicon disk array photonic sensors. The dots are the simulation points and the dashed lines are the best fits. b. The simulated SNR of the non-Hermitian plasmonic-photonic hybrid with a detector noise of 80 dB. The dashed lines are the fitting results of the SNR over the Q-factor for the non-Hermitian design based on Eq.~\ref{eqn:det1}. The two dashed boxes highlight the design at the EP and the design with the best SNR. We chose the latter as our sample for experimental characterization.}
    \label{fig3}
\end{figure}

Fig.~\ref{fig3}a compares our plasmonic-photonic hybrid with plasmonic and photonic systems. The plasmonic sensor considered here is a cubic array composed of a pair of Al nano-ellipsoids with 80 nm and 160 nm long axes separated by a 10 nm gap. The dielectric sensor is a hexagonal array of 60 nm thick silicon cylinders with a radius of 100 nm on a silica substrate. To calculate sensitivity for these structures, we modeled the analyte interaction by a 5 nm-thick dielectric layer whose index is set higher for higher concentrations of analyte. This layer wraps the ellipsoid-pair in our plasmonic system, sandwiches the silicon and silica layers in our hybrid system, and sits on top of silicon cylinders in our photonic system. The Q-factors of the plasmonic and photonic systems are varied by varying the period of the array. Varying spacer thickness from 50 nm to 120 nm in our hybrid case tuned its Q-factor.

The sensitivity vs. Q-factor plot for the three systems is shown in Fig.~\ref{fig3}a. Using Eq.~\ref{eqn:fitting}, we get a good fit for the observed relationship between the sensitivity and the Q-factor of our non-Hermitian system. All three systems show a trade-off between sensitivity and Q-factor \cite{li2023balancing}. However, there is a wide range of Q-factors in which our non-Hermitian sensor shows significantly higher sensitivity. Giving this high sensitivity up by a little to gain in Q-factor allows the non-Hermitian sensor to significantly boost its detectivity. Note that high detectivity requires both high Q-factor and high sensitivity. Thus, the high sensitivity over a wide range of Q-factors is the key to achieving high detectivity and off-EP non-Hermitian sensors are promising.

We simulated the detectivity of our non-Hermitian structure with various spacer thicknesses by adding an 80 dB detector noise to the reflectance spectra. For different values of $\alpha$, the net noise at the detector was calculated from our noise model (Fig.~\ref{fig1}c) and added to the spectra. The SNR vs. Q-factor is plotted for our non-Hermitian system in Fig.~\ref{fig3}b. When $\alpha>0$, the highest SNR operation corresponds to an off-EP condition. In our simulations, this optimum condition corresponded to the spacer thickness $h'=$ 60 nm. We chose this design for sample fabrication and experimental characterization.

\begin{figure}[htpb]
    \centering
    \includegraphics[width=1\textwidth]{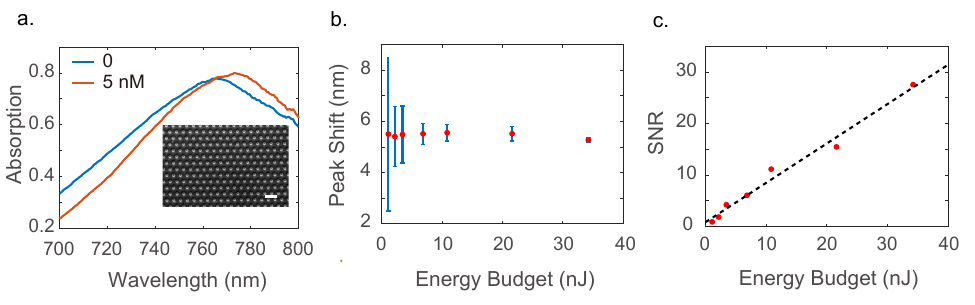}
    \caption{a. Measured absorption spectra of the non-Hermitian plasmonic-photonic sample with and without exposure to the anti-mouse IgG solution. The inset shows the SEM image of the as-fabricated sample with a spacer thickness $h'=$ 60 nm. The scale bar represents 1 \textmu m. b. The peak absorption wavelength shift upon analyte exposure evaluated from the Gaussian fits to the measured spectra at various incident optical energy budgets. The error bars are the standard deviations. c. The SNR versus the optical energy budget. The SNR is calculated from the error bars of b. The trend line shows the expected linear dependence with a slope of 0.765 $/$nJ.}
    \label{fig4}
\end{figure}

The scanning electron microscope image of the fabricated non-Hermitian sample is shown in the inset of Fig.~\ref{fig4}a. The absorption spectra of this device are shown in the same figure for no analyte and 5 nM analyte cases. The analyte here is an anti-mouse IgG protein. The functionalization of the non-Hermitian metasurface, the preparation of the anti-mouse IgG solution, and other details of biosensing are given in the methods section at the end. We used a microscope-based energy-momentum mapping setup to measure the reflectance and absorption spectra of our samples \cite{yang2022non}.  The resonance peak wavelength was extracted by fitting the measured spectrum with a Gaussian curve. From the shift in peak wavelength, the sensitivity of the sensor to the anti-mouse IgG protein is estimated to be about 1.23 nm$/$nM. This sensitivity is close to the highest reported for this protein in the literature \cite{cheng2020high}.

Next, we characterized the detectivity of this sensor by varying the optical energy budget. We held all the characterization parameters constant while changing only the illumination light intensity using a set of neutral density (ND) filters. We acquired the absorption spectrum on the sample ten times for each ND filter setting. Then, running Gaussian fits on the absorption spectra to estimate the peak shift, we assessed the standard deviation or error in the peak shift estimation. Fig.~\ref{fig4}b plots the resonance peak shift for 5 nM analyte concentration estimated at various incident optical energy budgets. The energy budget for each spectrum acquisition was obtained by measuring the incident optical power at the detector in the spectral range of 750 to 800 nm and multiplying it with the integration time of the detector. The error bars in Fig.~\ref{fig4}b represent the standard deviation for each measurement set.  Fig.~\ref{fig4}c plots the SNR of the device as a function of the energy budget. With the increase in the energy budget, the SNR increases correspondingly, and the trend line shows that the SNR is linearly proportional to the energy budget. The SNR grows linearly with the energy budget at the rate of 0.765 per nJ. Thus, the detectivity of the non-Hermitian sensor is 0.765 per nJ. This high detectivity simply means that about 1.3 nJ of a light pulse is sufficient to detect anti-mouse IgG protein with very high sensitivity.

\section{Conclusions}
Non-Hermitian sensors designed to work at EP are not the best choice in practice, especially when the energy budget is limited as on mobile platforms. Here, we showed that an off-EP non-Hermitian sensor is promising for real-world applications where both high sensitivity and high detectivity are required.  In a passive PT-symmetric system comprising a plasmonic-photonic hybrid, we showed that the detectivity peaks off-EP where the sensitivity can still be very high. We demonstrated a non-Hermitian nanophotonic sensor for sensing anti-mouse IgG protein and measured its sensitivity and detectivity. We reported one of the highest values of sensitivity reported in the literature for sensing anti-mouse IgG protein while needing a small minimum optical budget of 1.3 nJ. This work highlighted that non-Hermitian plasmonic-photonic hybrids could be the best class of nanophotonic sensors for real-world applications.

\begin{acknowledgement}
This work was supported by the Army Research Organization grant W911NF2120031.
\end{acknowledgement}




\section{Methods}
\noindent
\textbf{Noise simulations}:

Noise simulations in Fig.~\ref{fig1}d, and \ref{fig3}b were carried out by adding 80 dB detector noise (white Gaussian) to the resonant spectra, and running a Gaussian fit routine to extract the peak position. The fit routine was run 10000 times to get the standard deviation data. The signal-to-noise ratio (SNR) was calculated as the peak shift under the zero-noise condition divided by the standard deviation.

\noindent
\textbf{Electromagnetic simulations}:

Full-wave electromagnetic simulations were performed using a commercial finite-difference-time-domain solver (Lumerical). Si, $\rm SiO_2$, and Al optical constants were obtained from Palik~\cite{palik1998handbook} and fitted with Drude-Lorentz models. The simulations were carried out on a single period of the array with periodic boundary conditions. The light source was a plane wave source set 600 nm above the top of the metasurface structure. The frequency-domain field and power monitor planes were set 100 nm above the light source. Both the light source and the monitor were set to cover the entire simulation region. The mesh was uniform with x, y, and z grid spacing of 10 nm, 10 nm, and 5 nm, respectively.

\noindent
\textbf{Sample fabrication}: 

Planar nanofabrication was used for fabricating the metasurface. At first, a 120 nm thick aluminum layer was sputtered (AJA ATC Orion sputter system) on a fused silica substrate (0.5 mm thick, MTI Corp.). Next, layers of SiO$_{2}$ (60~nm), and Si (120~nm) were deposited using plasma-enhanced chemical vapor deposition (PlasmaTherm Versaline). A layer of e-beam resist (PMMA A4) and E-spacer was coated on the surface. Then, patterning was performed using standard e-beam lithography (Elionix ELS-G100). Next, a 20 nm-thick Al$_{2}$O$_{3}$ was deposited as an etch mask using an e-beam evaporator. Following a liftoff in acetone, the nanocylinders are formed using a subsequent reactive ion etch (Oxford, Plasmalab System 100/ICP 180) in a mixture of C$_{4}$F$_{8}$, SF$_{6}$, CF$_{4}$, and O$_{2}$ at flow rates of 50, 5, 25, and 2 sccm, respectively. The capacitively- and inductively-coupled RF powers were maintained at 75 and 200 W, respectively.

\noindent
\textbf{Functionalization and biosensing}:

We use a 0.1 M solution of 8-Hydroxyoctanoic acid (AmBeed, A200175) in ethanol to functionalize our non-Hermitian sample. We let the sample sit in this solution for 12 hours at 4$^{\circ}$C. Then we soak the sample in 2-(N-morpholino) ethanesulfonic acid (MES) buffer at a pH of 6.5 for 35 minutes. The MES buffer contained 0.4 M EDC or 1-ethyl-3-(3-dimethylaminopropyl) carbodiimide hydrochloride (Thermo Fisher) and 0.1 M NHS or N-Hydroxysuccinimide (Sigma Aldrich). After drying the sample, we incubated the device with 100 \textmu g/mL anti-CD63 antibodies (Ancell, 215-820) for one hour at room temperature. The surfaces were subsequently blocked with 5$\%$ Bovine Serum Albumin or BSA (Sigma Aldrich, A8531) in phosphate buffer solution or PBS (Thermo Fisher, 10010023) for 30 minutes at room temperature. After rinsing with PBS, we immersed our device in anti-mouse IgG (Sigma Aldrich, B7264) of a particular concentration at 4$^{\circ}$C for 12 hours. The original anti-mouse IgG solution is diluted with PBS buffer to solutions of concentration 5 nM. Then, after drying, the sample is subjected to optical characterization. After characterization, the same device is reused for sensing a different concentration of anti-mouse IgG. Before reusing, the device is rinsed to clean off all the added chemicals using SC-1 or a combination of ammonium hydroxide, hydrogen peroxide, and water in a volume ratio of 1:1:5 at room temperature for 90 minutes. The device is then thoroughly rinsed in DI water before reusing.

\noindent
\textbf{Spectra measurement}:

We use a Fourier-space imaging or energy-momentum imaging setup to measure the spectrum in the normal direction. This setup allows for single-shot measurement of angle-dependent reflection ($Ref$) and transmittance ($Tr$) spectra on small-area samples. Absorption ($Abs$) is calculated as $Abs=1-Tr-Ref$. Inserting a Bertrand lens into a standard imaging spectrophotometer allows projecting the Fourier space onto the imaging device. More details of the setup may be found in our previous work \cite{yang2022non}.

\noindent
\textbf{Optical energy budget}:

The optical energy budget for measurement is obtained by multiplying the integration time of the detector (10 s) with the optical power at the detector when the sample is replaced by a mirror. This power is extracted from the total photocurrent recorded by the power meter (ThorLabs, S121C). The spectral shape of the incident light from the quartz halogen bulb (Nikon, 12 V, 100 W, 2000 Hrs), and the transmittance spectra of the short-pass and long-pass filters (Thorlabs FES0800, FESL0700) are used in the estimation of the optical power.

\bibliography{references}

\end{document}